\documentclass[%
 reprint,
superscriptaddress,
 amsmath,amssymb,
 aps,
 showkeys
floatfix,
]{revtex4-1}

\usepackage{graphicx}
\usepackage{dcolumn}
\usepackage{bm}
\usepackage{algorithm}
\usepackage{algpseudocode}

\begin{document}


\title{An adaptive preconditioning scheme for the self-consistent field iteration and generalized stacking-fault energy calculations}





\author{Sitong Zhang}
\affiliation{Center for High Pressure Science, State Key Laboratory of Metastable Materials Science and Technology, Yanshan University, Qinhuangdao 066004, China.}

\author{Xingyu Gao}
\email{gao\_xingyu@iapcm.ac.cn}
\affiliation{Laboratory of Computational Physics, Institute of Applied Physics and Computational Mathematics, Fenghao East Road 2, Beijing 100094, People’s Republic of China.}

\author{Haifeng Song}
\affiliation{Laboratory of Computational Physics, Institute of Applied Physics and Computational Mathematics, Fenghao East Road 2, Beijing 100094, People’s Republic of China.}

\author{Bin Wen}%
\email{wenbin@ysu.edu.cn}
\affiliation{Center for High Pressure Science, State Key Laboratory of Metastable Materials Science and Technology, Yanshan University, Qinhuangdao 066004, China.}


\begin{abstract}
\begin{description}	
\item[Abstract]
The generalized stacking-fault energy (GSFE) is the fundamental but key parameter for the plastic deformation of materials. We perform first-principles calculations by full-potential linearized augmented planewave (FLAPW) method to evaluate the GSFE based on the single-shift and triple-shift supercell models. Different degrees of defects are introduced in the two models, thereby affecting the convergence of the self-consistent field (SCF) iterations. We present an adaptive preconditioning scheme which can identify the long-wavelength divergence behavior of the Jacobian during the SCF iteration and automatically switch on the Kerker preconditioning to accelerate the convergence. We implement this algorithm in Elk-7.2.42 package and calculate the GSFE curves for Al, Cu, and Si (111) plane $\langle\bar{1}\bar{1}2\rangle$ direction. 
We found that the single-shift and triple-shift supercell models have equivalent calculation accuracy and are within the experimental data uncertainty. 
For computational efficiency, the triple-shift supercell model is preferable due to its better convergence, exhibiting lower degree of defect compared to the single-shift supercell model.

\end{description}
\end{abstract}

\maketitle


\section{\label{sec:level1}INTRODUCTION }\label{sec:1}
In the field of materials science, defects in materials, including point defects, line defects, planar defects, bulk defects \cite{16,17,18,19}, significantly influence the mechanical, electronic, optical, and thermal properties of materials, thereby impacting their performance and applications \cite{ehrhart1992properties,siegel1981atomic,arrigoni2021spinney,davidsson2021adaq}. Understanding and characterizing these defects is crucial for studying material microstructure evolution.
In particular, stacking-fault is an importment kind of planar defect caused by the misalignment of atomic layers in crystals \cite{hutchinson1977creep,gilman1952surface}, and the stacking-fault energy (SFE) refers to the energy change when stacking-fault is formed in the crystal, which directly affects the mechanical properties of materials \cite{15,21}.
It is shown that metal materials with low SFE tend to slip mechanism, while those with high SFE are inclined towards twinning mechanism \cite{meyers2001onset}.
Significantly, the experimentally measurable SFE typically represents the intrinsic SFE.
As an extension of SFE, the generalized stacking-fault energy (GSFE) has been proposed to reflect the SFE change of the entire surface, which can help gain a deeper understanding of plastic deformation and offer theoretical support for relevant experiments \cite{sun2022hardness}. It also reveals the energy barrier which is necessary to overcome in order to arrive at the intrinsic SFE.

For the calculation of GSFE, there are two factors need to consider: the stacking-fault construction and the energy calculation method.
The stacking-fault construction includes the direct supercell method, indirect supercell method, and axial interaction method \cite{20,22,23}. Among these, the direct supercell method is widely recognized for the ability of the atomic position relaxation. And two models are widely used for the direct supercell method in face-centered cubic (FCC) materials: the single-shift and triple-shift supercell models \cite{24}. It is worth noting that different types of defects are introduced to the two models.
In order to obtain accurate GSFE values, the Kohn-Sham density functional theory (DFT) is highly desirable \cite{1,2}. 
Full-potential linearized augmented planewave (FLAPW) is a high-precision DFT calculation method  \cite{7,ambrosch2006linear}, which has developed several computational software packages, such as WIEN2K \cite{blaha2001wien2k}, FLEUR \cite{fleurCode}, EXCITING \cite{gulans2014exciting}, ELK \cite{38}, etc. These softwares are widely regarded as benchmarks in assessing bulk properties,
including equations of state (EOS) \cite{13,otero2011gibbs2}, elastic constant \cite{panda2006determination}, and so on.
The application of the FLAPW method in defect calculations (like GSFE calculations) is relatively few \cite{hoshino2001full,vsob2000electronic}.
It is shown that high-precision computational results of FLAPW can be as a reference supporting the development of other material calculation methods (such as machine learning) \cite{freitas2022machine,dragoni2018achieving}.

In periodic boundary conditions, these defect models generally have large volumes, posing a computational challenge due to the substantial resources required.
Hence, optimizing computational efficiency becomes a crucial concern.
Different configurations may affect the computational efficiency on two aspects: atomic relaxation and electronic self-consistent field (SCF) iteration. 
At the same time, the defect states introduced by defects may cause "charge sloshing" during SCF iterations in large systems \cite{31,39}. 
The essence of charge sloshing is the long-wavelength divergence of the Jacobian caused by the complete screening effect. When the input charge changes, it will cause the output charge violently oscillate in the low-frequency part, resulting in bad convergence. Although it often appears in metals, similar phenomenon may also occur in some defective semiconductor systems.
However, prior judgment of whether a system has charge sloshing usually relies on experience.
And in some complex systems, such as metal-insulator contact systems, it is difficult to determine a priori whether there is charge sloshing \cite{yu2018piezoelectricity,36}.
Kerker preconditioning is considered an effective scheme for suppressing the long-wavelength divergence behavior of the Jacobian \cite{resta1977thomas,30}, and its effectiveness has been verified in some softwares \cite{32,33}. 
But imposing Kerker preconditioning on all systems may not achieve good convergences.
This raises the question: How to identify long-wavelength divergence behavior of the Jacobian during SCF iterations?

In this work, we found a good approximation of the inverse Jacobian in the Pulay subspace and constructed a posterior indicator to identify the long-wavelength divergence behavior of the Jacobian.   
Then, we developed an adaptive algorithm based on the indicator and the Kerker preconditioner in ELK codes.
Our algorithm can identify the charge sloshing and automatically switch to Kerker preconditioning, enabling efficient convergence of different stacking-fault models.
On this bases, we used Al, Cu, and Si as representative materials and evaluated the GSFE curve calculations from two aspects of accuracy and efficiency for (111) plane $\langle\bar{1}\bar{1}2\rangle$ direction with the different planar defect models: the single-shift and triple-shift supercell models, and explored the impact of different defects in stacking-fault models on calculations.

This paper is organized as follows: In Sec \ref{sec:2}, we will introduce fundamental models and computational methods. In Sec \ref{sec:3}, we will discuss the construction of a posterior indicator and the adaptive preconditioning scheme. In Sec \ref{sec:4}, we will present our results and analysis of the GSFE curves calculation. Concluding remarks will be presented in the last section.

\section{Basic models and calculation methods}\label{sec:2}
\subsection{Construction of stacking-fault model}
Vitek defined the SFE as \cite{21}:
\begin{equation}
	\begin{aligned}	
		\gamma=\frac{\Delta E}{s}.
	\end{aligned}
\end{equation}
Where $ \Delta E $ is the energy difference between the original structure and the stacking-fault structure, $s$ is the area of the slip plane.

We first focus on the modeling methods of the single-shift and triple-shift supercell models for GSFE curve calculations. 
In Al, Cu, and Si structures, the stacking sequence of the close-packed atomic layers along the z-$\langle111\rangle$ direction can be denoted by ABCABC.
Here, we define a 6-layer stacking configuration as the supercell unit block to mitigate the interaction impact between two stacking-faults along the z-axis direction. The $\langle{1}\bar{1}0\rangle$ and $\langle{1}0\bar{1}\rangle$ directions on the (111) plane are selected as the basis vectors to make the supercell unit block as small as possible to reduce computational cost. 

\begin{figure*}
	\includegraphics[width=0.8\textwidth]{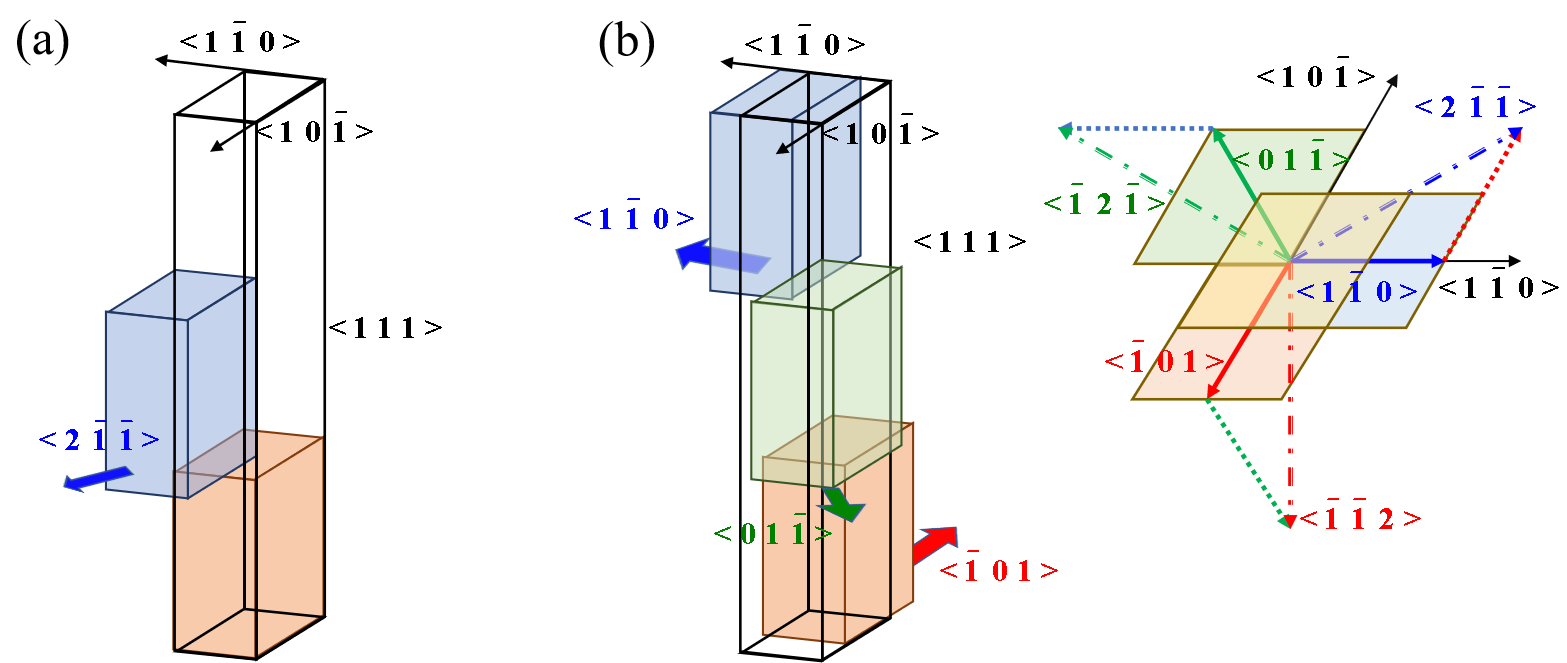}
	\caption{\label{fig:1}Schematic representation of the single-shift and triple-shift supercell models. (a)In the single-shift supercell model, the blue and red grids represent the 6-layer supercell units block, while the arrows indicate the slip directions of the grids. (b) Side and top views of the triple-shift supercell model, the yellow parallelogram represents the original position of the supercell. The blue, green, and red ones, respectively, indicate the upper, middle, and lower grids' displacements relative to the original position along the $\langle{1}\bar{1}0\rangle$, $\langle01\bar{1}\rangle$ and $\langle\bar{1}01\rangle$ directions (solid lines). The dotted-dashed lines represent the actual relative displacements between adjacent blocks, which are $\langle2\bar{1}\bar{1}\rangle$, $\langle\bar{1}2\bar{1}\rangle$, and $\langle\bar{1}\bar{1}2\rangle$.}
\end{figure*}

The single-shift supercell model consists of two interacting supercell unit blocks, where one remains stationary, and the other moves along the $\langle\bar{1}\bar{1}2\rangle$ direction. However, due to periodic boundary conditions, a relative displacement in the $\langle\bar{1}\bar{1}2\rangle$ direction between the blocks is induced. We add a 10 Å vacuum layer in the z-axis direction to eliminate the interaction between the blocks and their periodic images.
The triple-shift supercell model consists of three interacting supercell blocks. They simultaneously slips in the $\langle{1}\bar{1}0\rangle$, $\langle01\bar{1}\rangle$ and $\langle\bar{1}01\rangle$ directions, respectively. As a result, the displacements between adjacent blocks are $\langle2\bar{1}\bar{1}\rangle$, $\langle\bar{1}2\bar{1}\rangle$, and $\langle\bar{1}\bar{1}2\rangle$. The sum of these three displacement vectors is zero, no need to add a vacuum layer and preserving the periodic boundary conditions along the z-axis direction.
The modeling methods of the two models are shown in the Fig.~\ref{fig:1}.
We took 18 equidistant points to approximate the unit (111) plane $\langle\bar{1}\bar{1}2\rangle$ direction GSFE curve of Al, Cu, and Si (the (111) plane of Si includes the glide plane and the shuffle plane, and we only compute the glide plane). 

\subsection{FLAPW method}
We utilized ELK-7.2.42 packages. It is based on the FLAPW method, which divides the space of the crystal into two regions: muffin-tin spheres and interstitial regions. The wavefunctions are expanded using a combination of atomic orbitals in the muffin-tin region and planewave in the interstitial region. The wave functions $ \psi^\sigma_{\nu\textbf{k}} $
for spin-$ \sigma $ valence orbital of band index $ \nu $ with Bloch vector 
\textbf{k} is:
\begin{equation}
	\psi^\sigma_{\nu\textbf{k}}(\textbf{r})=\sum_{\textbf{k}+\textbf{G}}c_{\nu\textbf{k}}^{\textbf{G}\sigma}\varphi^\sigma_{\textbf{{G}}}(\textbf{k}, \textbf{r}).
	\label{eq:aa}
\end{equation}
Where $c_{\nu \textbf{k}}^{\textbf{G}\sigma}$ is the expansion coefficients of the wavefunction, $ \textbf{G} $ are all reciprocal lattice vectors dual to the lattice vectors defining the periodic domain up to the  $ \textbf{G}_{\rm{max}} $, and the basis functions $ \varphi^\sigma_{\textbf{{G}}}(\textbf{k}, \textbf{r}) $ can be written as:
\begin{equation}
	\varphi^\sigma_{\textbf{{G}}}(\textbf{k}, \textbf{r}) = \begin{cases}
		\sum_{jL\alpha}A^{\sigma \alpha \textbf{G}}_{jL}(\textbf{\textit{k}})u^{\sigma \alpha}_{jl}(r_\alpha)Y_L(\hat{\bf{r}}_\alpha)& \rm{muffin-tin} \\	
		\frac{1}{\sqrt{\Omega}}\exp({i}(\textbf{k}+\textbf{G})\cdot\textbf{{r}})& \rm{interstitial}. \\
	\end{cases}	
	\label{eq:a}
\end{equation}
Where $ L $ abbreviates the quantum numbers $ l $ and $ m $, $ \alpha $ is the muffin-tin sphere, $ Y_L(\hat{\bf{r}}) $ is the spherical harmonics, $ A^{\sigma \alpha \textbf{G}}_{jL}(\textbf{\textit{k}}) $ is the matching coefficient, $ j $ is the order of the basis function, for $ j = 0 $ and 1, $ u^{\sigma \alpha}_{jl}(r_\alpha) $ is the solution of the radial Schrödinger equation at a linearization energy and its energy derivative evaluated at this same energy. In the interstitial region, the function is the planewave.
As the basis sets in different regions of the crystal are different, it results in different representations of the density and potential in the two regions:
\begin{equation}
	\rho(\textbf{{r}}) = \begin{cases}
		\sum_{L}\rho^\alpha_L({r}_\alpha)Y_L({\hat{\textbf{r}}}_\alpha)	 & \rm{muffin-tin}   \\	
		\sum_{{\textbf{k}}+{\textbf{G}}}\rho^{\rm{I}}_{{({\textbf{k}}+{\textbf{G}})}}\exp({i}({\textbf{k}}+{\textbf{G}})\cdot\textbf{{r}})	 & \rm{interstitial} ;\\
	\end{cases}	
\end{equation}
\begin{equation}
	V(\textbf{{r}}) = \begin{cases}
		\sum_{L}V^{\rm{MT}}_LY_L({\hat{\textbf{r}}}_\alpha)	 & \rm{muffin-tin}   \\	
		\sum_{{\textbf{k}}+{\textbf{G}}}V^{\rm{I}}_{({\textbf{k}}+{\textbf{G}})}\exp({i}({\textbf{k}}+{\textbf{G}})\cdot\textbf{{r}})	 & \rm{interstitial}.\\
	\end{cases}	
\end{equation}
Based on the above, an eigenvalue problem can be derived:
\begin{equation}
	\hat{H}	\psi^\sigma_{\nu\textbf{k}}(\textbf{r})=\varepsilon_{\nu\textbf{k}}\psi^\sigma_{\nu\textbf{k}}(\textbf{r})
	\label{eq:bb}
\end{equation}
where Hamiltonian $ \hat{H} $ is a sum of corresponding terms:
\begin{equation}
	\hat{H}=\hat{T}_0+\hat{V}_{\rm ext}+\hat{V}_{\rm H}+\hat{V}_{\rm xc}.
\end{equation}
$ {T}_0 $ is kinetic energy, $  {V}_{\rm ext} $ is external-potential, $ {V}_{\rm H} $ is Hartree potential, $ {V}_{\rm xc} $ is exchange-correlation potential. 
Of which, $ {V}_{\rm xc} $ and  $ {V}_{\rm H} $ are the local potentials and explicitly density dependent, and the charge density depends on the solution of the Eq.\eqref{eq:bb}, leading to the nonlinearity of the Kohn-Sham equation. One need to resort to a fixed-point iteration scheme:
\begin{equation}
	\begin{aligned}	
		\textbf{n}_{m+1}=\textbf{Q}(\textbf{n}_{m})
	\end{aligned}
\end{equation}
The residual $\textbf{f}(\textbf{n}_m)$ is defined as: $\textbf{f}(\textbf{n}_m)=\textbf{Q}({\textbf{n}_m})-\textbf{n}_m.$
Based on this idea, the simple mixing scheme for the vector $ \textbf{n}_{m} $ (represents charge density $ \rho_m $ or potential $ {v}_m $, in this work, we employ potential mixing) can be derived:
\begin{equation}
	\begin{aligned}	
		\textbf{n}_{m+1}
		=\textbf{n}_m +P\textbf{f}(\textbf{n}_m) ,
		\label{eq:bl}
	\end{aligned}
\end{equation}
where $ P $ is the preconditioner.
Through SCF iterations and continuously updating the charge density and potential until the desired convergence accuracy is achieved, the ground state-energy can be obtained. From this, one can further derive various physical properties.

\subsection{GSFE curve calculation test}
\begin{figure}
	\includegraphics[width=0.45\textwidth]{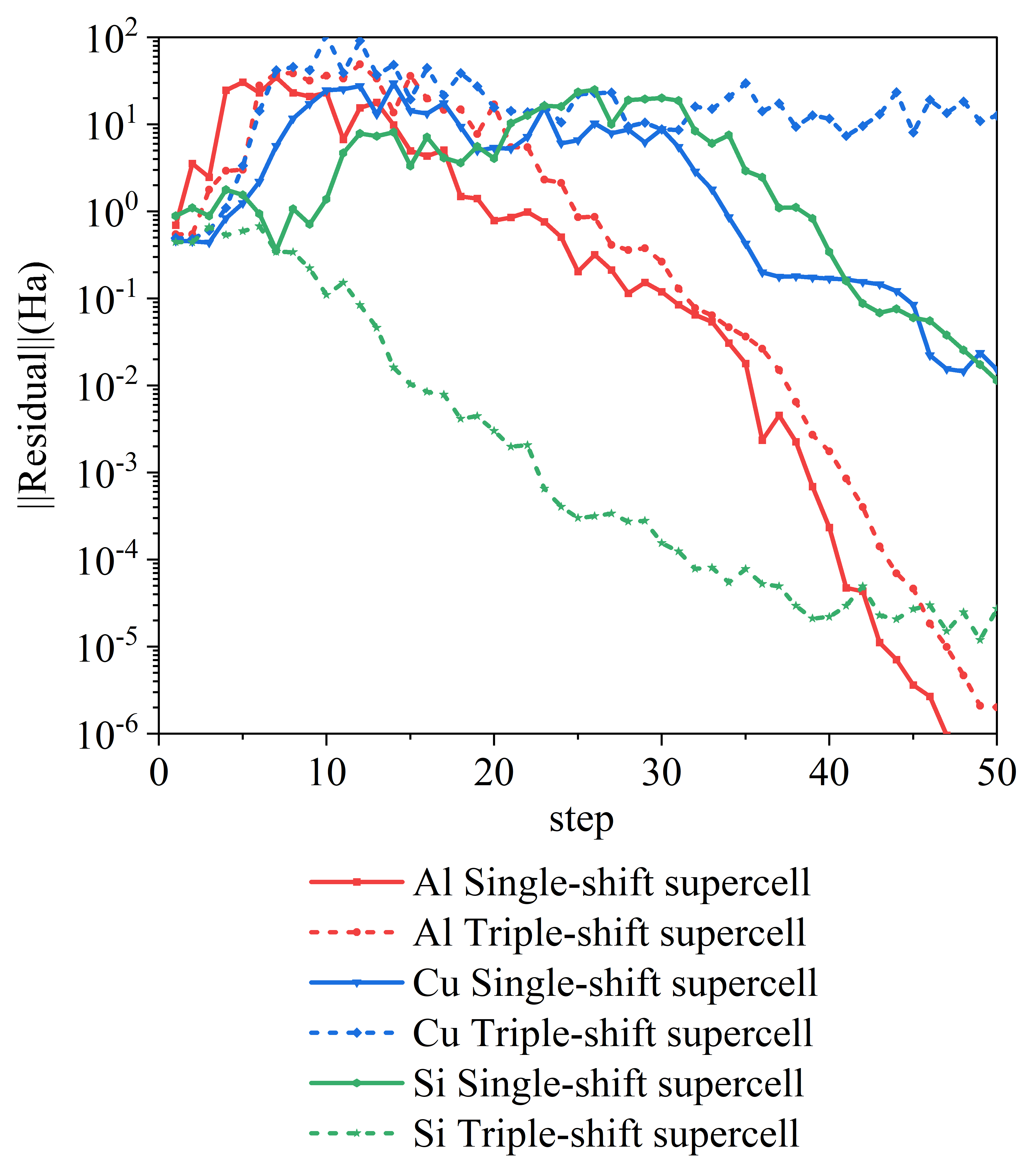}
	\caption{\label{fig:2} The convergence of residual for Al, Cu and Si of the single-shift and triple-shift supercell models by using the original Pulay mixing scheme.}
\end{figure}
 We employed the FLAPW combined with the local orbital basis set in ELK, the muffin-tin radii were set to $R_{\rm MT}^{\rm Al}=2$ bohr, $R_{\rm MT}^{\rm Cu}=2.2$ bohr, and $R_{\rm MT}^{\rm Si}=2$ bohr for Al, Cu, and Si, respectively. The cut-off radii for these systems were all set to $R^{\rm MT}_{\rm mix}K_{\rm max}=10$. The Brillouin zone sampling was set as $11\times11\times1$ for k-points. We adopt the generalized gradient approximation in the Perdew–Burke–Ernzerhof parametrization for the exchange-correlation functional \cite{37}. 
We employed the original Pulay mixing scheme \cite{65}, 
and set the convergence criterion as energy between two consecutive iteration steps being smaller than $10^{-6}$ Hartree. All atoms relax along the z-axis direction during the calculation, and atomic forces 
were converged to less than $0.001$ Hatree/bohr.

To calculate the GSFE curve in the $\langle\bar{1}\bar{1}2\rangle$ direction, we took 18 equidistant points along the unit $\langle\bar{1}\bar{1}2\rangle$ direction, and adopted the  1/18$\langle\bar{1}\bar{1}2\rangle$ point as an example to show the convergence of the single-shift and triple-shift supercell models calculations in Fig.~\ref{fig:2}. 
It is shown that only the single-shift supercell model of Al converges in 50 steps; others do not converge in 50 steps. Obviously, calculating the GSFE curve under current scheme is inefficient.
\section{Posterior Indicators and the Adaptive Preconditioning scheme}\label{sec:3}
Previous studies have reported similar convergence problems, i.e., the long-wavelength divergence behavior of the Jacobian \cite{31,39}.
We want to identify the long-wavelength divergence behavior of the Jacobian in computations and accelerate convergence. 
In this section, we first present the derivation of a posteriori indicator to monitor the eigenvalue behavior of the Jacobian. Then, we briefly describe the implementation of Kerker preconditioning under the FLAPW basis. And finally, we combine them to obtain an adaptive preconditioning scheme.

\subsection{A Posterior Indicator}
Firstly, we define the Jacobian as:
\begin{equation}
	J=-\frac{d\textbf{f}(\textbf{n}_m)}{d\textbf{n}_m}.
	\label{eq:ed}
\end{equation}
The residual propagation of Eq.\eqref{eq:bl} is given
by:
\begin{equation}
	\textbf{f}(\textbf{n}_{m+1}) \approx(I- JP)\textbf{f}(\textbf{n}_m) .
	\label{eq:edd}
\end{equation}
The convergence condition can be written as:
\begin{equation}
	||I-JP||<1.
	\label{eq:ef}
\end{equation}
If there are small eigenvalues in  $(JP)^{-1} $,  the conditions of Eq.\eqref{eq:ef} are difficult to satisfy, which implies the long-wavelength divergence behavior of the Jacobian. But it is hard to calculate the eigenvalues of $(JP)^{-1} $. Practically, we can use the subspace of Pulay scheme to approximate $(JP)^{-1} $.

In the Broyden's second method, a sequence of low-rank modifications are made to modify the initial guess of the inverse Jacobian, 
the recursive formula can be derived from the following constrained optimization problem \cite{44,45}:
\begin{equation}
	\begin{aligned}	
		{\rm min}_H\quad \frac{1}{2}||H-H_{m-1}||^2_F\\
		{\rm such\quad that}\quad 	H_mF_{m-1} = -N_{m-1}
		\label{eq:qq}
	\end{aligned}
\end{equation}
where $ H_{m-1} $ is the approximation to the inverse Jacobian
in the $ (m-1) $th Broyden update. $ N_{m-1} $ and $ F_{m-1} $ are:
\begin{equation}
	\begin{aligned}	 
		N_{m-1} =(\delta \textbf{n}_{m-1},\dots,\delta \textbf{n}_{m-\ell+1});
	\end{aligned}
	\label{eq:c}
\end{equation}
\begin{equation}
	\begin{aligned}	 
		F_{m-1} =(\delta \textbf{f}_{m-1},\dots,\delta \textbf{f}_{m-\ell+1});
	\end{aligned}
	\label{eq:d}
\end{equation}
where $\ell$ is the size of the subspace.
Eq.\eqref{eq:qq} has an analytical solution as \cite{36}:
\begin{equation}
	\small
	H_m = H_{m-1}-(N_{m-1} + H_{m-1}F_{m-1})( F_{m-1}^TF_{m-1})^{-1}F^T_{m-1}.\quad
	\label{eq:bd}
\end{equation}
And we can replace the $ H_{m-1} $
in Eq.\eqref{eq:bd} to the initial guess $ H_1 $ as the inverse of Jacobian:
\begin{equation}
	\small
	H_m = H_1-(N_{m-1} + H_1F_{m-1})( F_{m-1}^TF_{m-1})^{-1}F^T_{m-1}.\quad
	\label{eq:bdd}
\end{equation}
Then we follow the quasi Newton approach to solve the fixed-point iteration, this leads to the Pulay mixing scheme, which is also the method used in Elk codes:
\begin{equation}
	\begin{aligned}	
		\textbf{n}_{m+1}&=\textbf{n}_m+	H_m\textbf{f}(\textbf{n}_m)\\
		& = \textbf{n}_m +H_1\textbf{f}(\textbf{n}_m)-(N_{m-1} + H_1F_{m-1})\\
		&\quad\times( F_{m-1}^T F_{m-1})^{-1}F^T_{m-1}\textbf{f}(\textbf{n}_m),
		\label{eq:q7}
	\end{aligned}
\end{equation}
The $ H_{m} $ of Eq.\eqref{eq:bdd} satisfies the constraint condition:
\begin{equation}
	H_mF_{m-1} = -N_{m-1}
	\label{eq:egg}
\end{equation}
and for the Jacobian $ J $, it also satisfies:
\begin{equation}
	J^{-1}F_{m-1} \approx -N_{m-1}.
	\label{eq:egp}
\end{equation}
From Eq.\eqref{eq:egg} and Eq.\eqref{eq:egp}, we can calculate the eigenvalues of $ P^{-1}H_m $ replacing the eigenvalues of $ (JP)^{-1} $, because $ H_m $ provides an effective approximation to $ J $ within the subspace defined by $ F_m $.
The eigenvalues of  $ P^{-1}H_m $ can be obtained by solving a generalized eigenvalue problem:
\begin{equation}
	F_{m-1}^TH_mF_{m-1}\textbf{u}_i =\lambda_i F_{m-1}^TPF_{m-1}\textbf{u}_i,
	\label{eq:eh}
\end{equation}
Eq.\eqref{eq:eh} can be shift as:
\begin{equation}
	F_{m-1}^T(H_m-P)F_{m-1}\textbf{u}_i =(\lambda_i-1) F_{m-1}^TPF_{m-1}\textbf{u}_i.\quad
	\label{eq:ei}
\end{equation}
We only need to calculate the eigenvalues of the left-hand side of the Eq.\eqref{eq:ei}, select the smallest eigenvalue as a posterior indicator. When the indicator is too small, it signifies the long-wavelength divergence behavior of the Jacobian. Note that solving Eq.\eqref{eq:ei} requires a little computational overhead because $ (H_m-P)F_{m-1} $ has been calculated in Pulay's update:
\begin{equation}
	(H_m-P)F_{m-1} = -(N_{m-1} + H_1F_{m-1})
	\label{eq:e77}
\end{equation}
 and the size of the subspace $\ell$ is generally smaller than 50. The details of relevant derivation can be found in Ref \cite{36}.

\begin{figure}
	\includegraphics[width=0.45\textwidth]{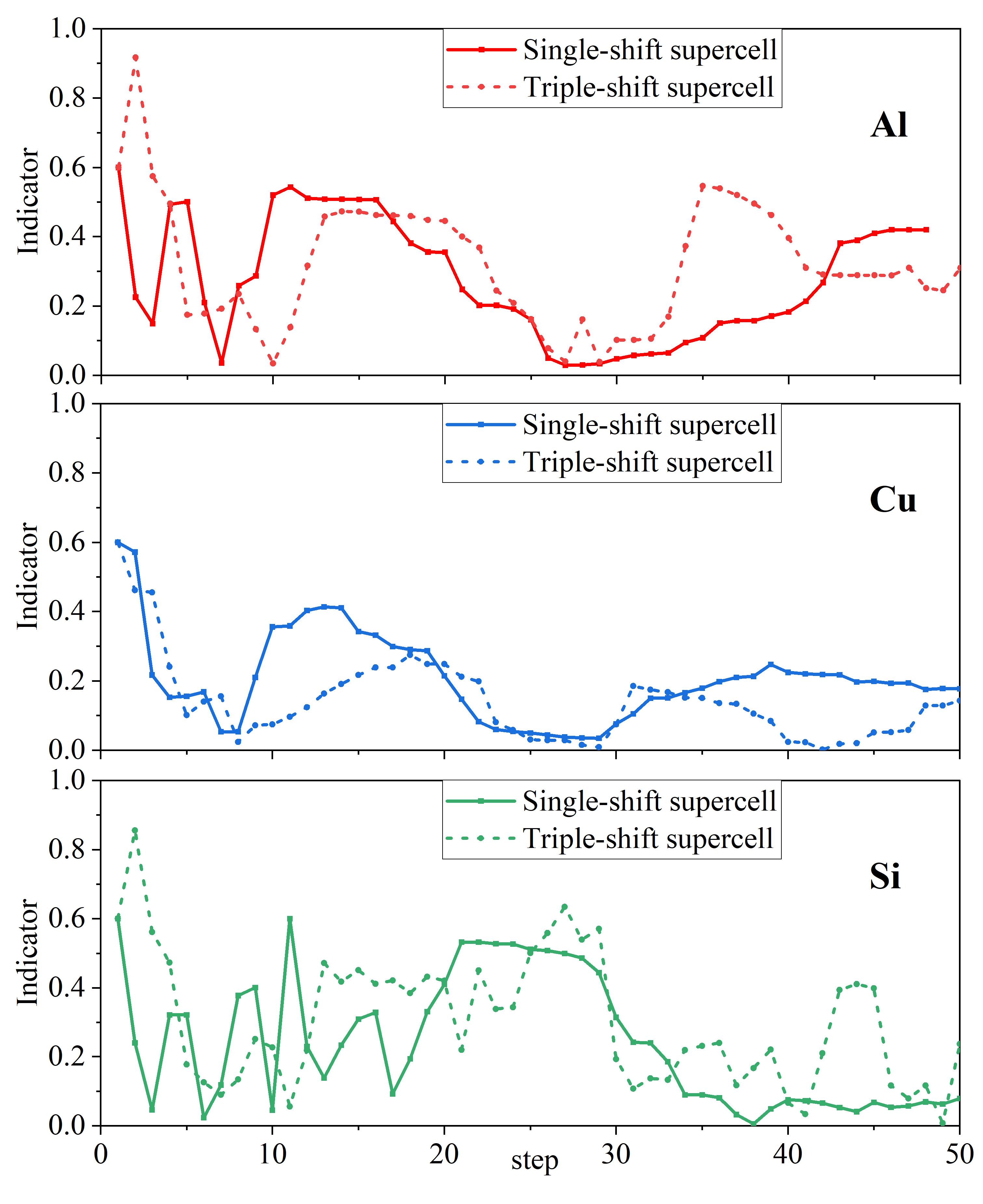}
	\caption{\label{fig:3} The indicator for Al, Cu and Si of the single-shift and triple-shift supercell models by using ELK.}
\end{figure}

In the general scheme (also in our tests), $ H_1 $ of Eq.\eqref{eq:bdd} is set to $ \alpha I $, where $ \alpha $ is a scalar parameter usually set to 0.4.
The indicators of Al, Cu, and Si systems for the single-shift and triple-shift supercell models tests are shown in Fig.~\ref{fig:3}. We found that they all have small indicators during the SCF iterations. It means that the convergence condition of Eq.\eqref{eq:ef} cannot be satisfied, and the residual may be magnified in the corresponding step, leading to the SCF iteration is hard to converge.
\subsection{Kerker preconditioning in the FLAPW method}
When we find the long-wavelength divergence behavior of the Jacobian through a posterior indicator, an effective method to suppress the long-wavelength divergence behavior of the Jacobian is the Kerker preconditioner. 
The Kerker preconditioner is based on the Thomas-Fermi screening model. The dielectric function of the Thomas-Fermi screening model in reciprocal space can be expressed as:
\begin{equation}
	\varepsilon(\textbf{G})=\frac{\textbf{G}^2+k_{\rm TF}^2}{\textbf{G}^2},
	\label{eq:f}
\end{equation}
where $ k_{\rm TF} $ is  the Thomas-Fermi vector \cite{36}. The Kerker preconditioner can be derived by inverting the dielectric matrix \cite{46,47}, and taking the Kerker preconditioner into simple mixing scheme can be expressed as:
\begin{equation}
	\begin{aligned}	
		\textbf{n}_{m+1}(\textbf{G})=\textbf{n}_m(\textbf{G}) +\alpha\frac{\textbf{G}^2}{\textbf{G}^2+k_{\rm TF}^2}\textbf{f}\left(\textbf{n}_m\left(\textbf{G}\right)\right) .
		\label{eq:i}
	\end{aligned}
\end{equation}
In the FLAPW method, due to the violent oscillation of the charge density near the nucleus, the Fourier transform is hard, so the Kerker preconditioner must be implemented in real space \cite{7}:
\begin{equation}
	\begin{aligned}	 
		\textbf{n}_{m+1}({\bf r}) = \textbf{n}_{m}({\bf r})+\alpha[1+\lambda^2(\nabla^2-\lambda^2)^{-1}]\textbf{f}\left(\textbf{n}_m\left(\textbf{r}\right)\right).
		\label{eq:j}
	\end{aligned}
\end{equation}
We replace the $ k_{\rm TF} $ by $ \lambda $, because Thomas-Fermi wave number of the homogeneous electron gas and the electronic properties of the actual system are hard to equal \cite{32}. Then, the key of the problem is the treatment of $ (\nabla^2-\lambda^2)^{-1} $,  which can be related to the solution of screened Poisson's equation \cite{47,49}:
\begin{equation}
	\begin{aligned}
	(\nabla^2 - \lambda^2)V(\textbf{r})= - 4\pi\rho(\textbf{r}).
	\label{eq:k}
 	\end{aligned}
\end{equation}
For the potential mixing, where  ${\rho}(\textbf{r})=-{v}_m(\textbf{r}) /4\pi$.
In order to solving the screened Poisson's equation in the FLAPW basis, we need to divide the solution into the muffin-tin and the interstitial regions, and firstly solve the screened Poisson's equation for the interstitial regions in reciprocal space, the solution of interstitial regions provides the boundary conditions for muffin-tin regions. Then, the solution of the muffin-tin regions can be solved by Green's function method. This is based on the pseudo-charge method proposed by Weinert \cite{48}.

For the solution of the interstitial region, we constructed the pseudo-charge  $ \bar{{\rho}}(\textbf{r}) $ that can be expanded in terms of the finite set of \textbf{G}-vectors.
The charge density in the interstitial region is extended to the entire space, and the charge density in the muffin-tin region is the original charge density subtracted from the extended interstitial charge density. We now replace the real muffin-tin charge density with a smooth density $ \tilde{{\rho}}^\alpha(\textbf{r}) $:
\begin{equation}
	\bar{{\rho}}(\textbf{r})={\rho}_{\rm I}(\textbf{r})+\sum_{\alpha}\tilde{{\rho}}^\alpha(\textbf{r}).
	\label{eq:n}
\end{equation}
Concurrently, the pseudo-charge must preserve equivalent multipole moments (the same as the real charge distribution) within the muffin-tin region. This implies that the multipole moments of smooth density $ \tilde{q}^{\alpha}_{lm} $ are determined by the difference between the muffin-tin and interstitial multipole moments:
\begin{equation}
	\begin{aligned}	 
		\tilde{q}^{\alpha}_{lm}=q^{\alpha,\rm{MT}}_{lm}-q_{lm}^{\rm I}.
		\label{eq:q}
	\end{aligned}
\end{equation}
Where the multipole moments for the muffin-tin $ q^{\alpha,\rm{MT}}_{lm} $ and interstitial regions $ q_{lm}^{\rm I} $ can be expressed as:
\begin{equation}
	\begin{aligned}	 
		q^{\alpha,\rm{MT}}_{lm}=\frac{{(2l+1)!!}}{\lambda^l}\int_{S_\alpha}Y_{lm}^*({\hat{\bf r}})i_l(\lambda r){\rho}{({\bf r})}d{\bf r};
		\label{eq:o}
	\end{aligned}
\end{equation}
\begin{equation}
	\begin{aligned}	
		q_{lm}^{\rm I}=&\frac{(2l+1)!!}{\lambda ^l}4\pi {\rm i}^l\sum_{\bf G}\rho^{\rm I}({\bf G})\exp({\rm i}{\bf G}\cdot{\bf r}_{\alpha})Y_{lm}^*
		(\hat{\bf G})\\
		&\times \int_0^{R_\alpha}i_l(\lambda r_\alpha)j_l(Gr_\alpha)r^2_\alpha dr_\alpha;\\
		\label{eq:p}
	\end{aligned}
\end{equation}
$ i_l(r) $ and $ k_l(r) $ are the modified spherical Bessel function of the first and second kind.
Then the pseudo-charge density in reciprocal space can be written as:
\begin{equation}
	\begin{aligned}	 
		\bar{\rho}{({\bf G})}=&{\rho}^I{({\bf G})}+\frac{4\pi}{\Omega}\exp({{\rm i}{\bf{G}\cdot\bf{R}}^\alpha})\sum_{l=0}^{\infty}\sum_{m=-l}^{l}\frac{{(-{\rm i})^l}}{(2l+1)!!}\\
		&\times\frac{\lambda ^{l+n+1}j_{l+n+1}(GR_\alpha)}{i_{l+n+1}(\lambda R_\alpha)G^{n+1}}Y_{lm}^*({\hat{\bf G}})\tilde{q}^\alpha_{lm}.
		\label{eq:s}
	\end{aligned}
\end{equation}
The pseudo-charge has a more rapidly converging Fourier expansion than the original charge density, and the $ V_I({\bf G}) $ in the interstitial can be get:
\begin{equation}
	\begin{aligned}	 
		V_I({\bf G})=\frac{4\pi}{G^2+\lambda}\bar{\rho}{({\bf G})}.
		\label{eq:t}
	\end{aligned}
\end{equation}

The potential of the interstitial region provides the value at the boundary of the muffin-tin sphere. The solution of potential within the muffin-tin sphere is a Dirichlet boundary-value problem, which can be solved by the Green's function method \cite{50}:
\begin{equation}
	\begin{aligned}	 
		V_{\rm MT}(\textbf{r})&=\int_{{\rm MT}_\alpha} G({\bf r},{\bf r}')\rho({\bf r}')d{\bf r}'-\frac{R_\alpha^2}{4\pi}\oint_{S_\alpha} V_I({ R_\alpha})\frac{\partial G}{\partial n'}d\Omega'\\
		&={4\pi\lambda}[k_l(\lambda r) \int_0^r\rho_{lm}(r')i_l(\lambda r')     {r'}^{2}dr'\\
		&+i_l(\lambda r) \int_r^{R_\alpha}{r'}^{2}{\rho_{lm}(r')}{k_l(\lambda r') }dr'\\
		&-i_l(\lambda r)\frac{k_l(\lambda R_\alpha)}{i_l(\lambda R_\alpha)}\int_0^{R_\alpha}\rho_{lm}(r')i_l(\lambda r')r'^{2}  dr'	]\\
		&+V^I_{lm}(R_\alpha)\left[\frac{i_l(\lambda r_\alpha)}{i_l(\lambda R_\alpha)}\right].
		\label{eq:u}
	\end{aligned}
\end{equation}

\begin{figure*}
	\includegraphics[width=0.7\textwidth]{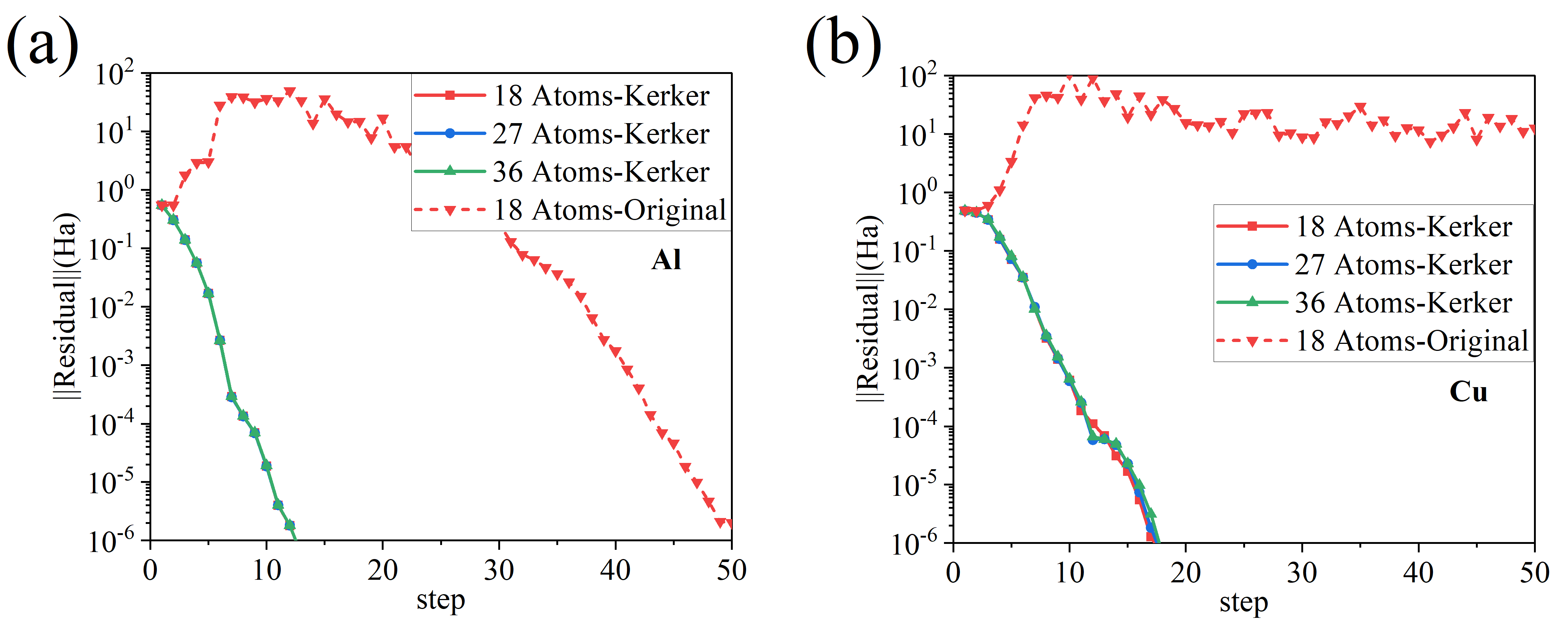}
	\centering
	\caption{\label{fig:4} (a)-(b)The convergence of residual for the triple-shift supercell model of Al, Cu with different thicknesses by using Kerker preconditioner respectively.}
\end{figure*}

One characteristic of Kerker preconditioner is that its convergence is not affected by an increase in the number of atoms on the long-axis for metal systems.
Based on the implementation of the Kerker preconditioner, we tested the convergence for the triple-shift supercell model of Al, Cu with different thicknesses on the z-axis. We set $\lambda_{\rm Al}=0.6$, $\lambda_{\rm Cu}=0.8$ \cite{32} 
, the convergence results are shown in Fig.~\ref{fig:4}.
We found that the convergence speed is significantly improved by using the Kerker preconditioner. At the same time, we found that the step of convergence almost unchanged for different atomic layer thicknesses. Obviously, the Kerker preconditioner successfully suppressed the long-wavelength divergence behavior of the Jacobian.
\subsection{The Adaptive Preconditioning scheme}
In the SCF iteration, we establish an initiation criterion for the Kerker preconditioner by setting a threshold of less than 0.2 for a posterior indicator. This leads to an adaptive preconditioning algorithm. The algorithm process is shown in Algorithm.~\ref{alg:ak}.
\begin{algorithm}[H]
	\caption{ An adaptive preconditioning algorithm}
	\label{alg:ak}
	\begin{algorithmic}[1]
		\State \textbf{Input:} $ \textbf{n}_0 $, $ \alpha $, $ \lambda $, tol;
		\Repeat\textbf{:}  $ m=1, 2, 3, 4 \dots $
		\State $\textbf{n}_{m+1}=\textbf{n}_m+H_m\textbf{f}(\textbf{n}_m)$, $H_1=\alpha I$;
		\If  {indicator < 0.2 }  ;
		\State  \textbf{Break}		
		\State  Initiate Kerker preconditioner, and reset $ m=1$		
		\Repeat\textbf{:}  $ m=1, 2, 3, 4 \dots $		
		\State $\textbf{n}_{m+1}=\textbf{n}_m+H_m\textbf{f}(\textbf{n}_m)$, $H_1=\alpha(\nabla^2-\lambda^2)^{-1}$;		
		\Until {$\sqrt{\frac{||\textbf{f}||^2}{\Omega}}<tol $} 
		\State \textbf{Output:} $ \textbf{n}_{m+1} $
		\EndIf
		\Until {$\sqrt{\frac{||\textbf{f}||^2}{\Omega}}<tol $} 
		\State \textbf{Output:} $ \textbf{n}_{m+1} $
	\end{algorithmic}
\end{algorithm}	
At the beginning of the program, we set $ H_1 $ to $\alpha I$. A posterior indicator is calculated in every SCF iteration. When the indicator is less than 0.2, break the current step, and the current subspace will also be discarded. Subsequently, a new SCF iteration will be started with the Kerker preconditioner initiating, continue iteration until convergence accuracy is reached.

\begin{figure*}
	\includegraphics[width=0.9\textwidth]{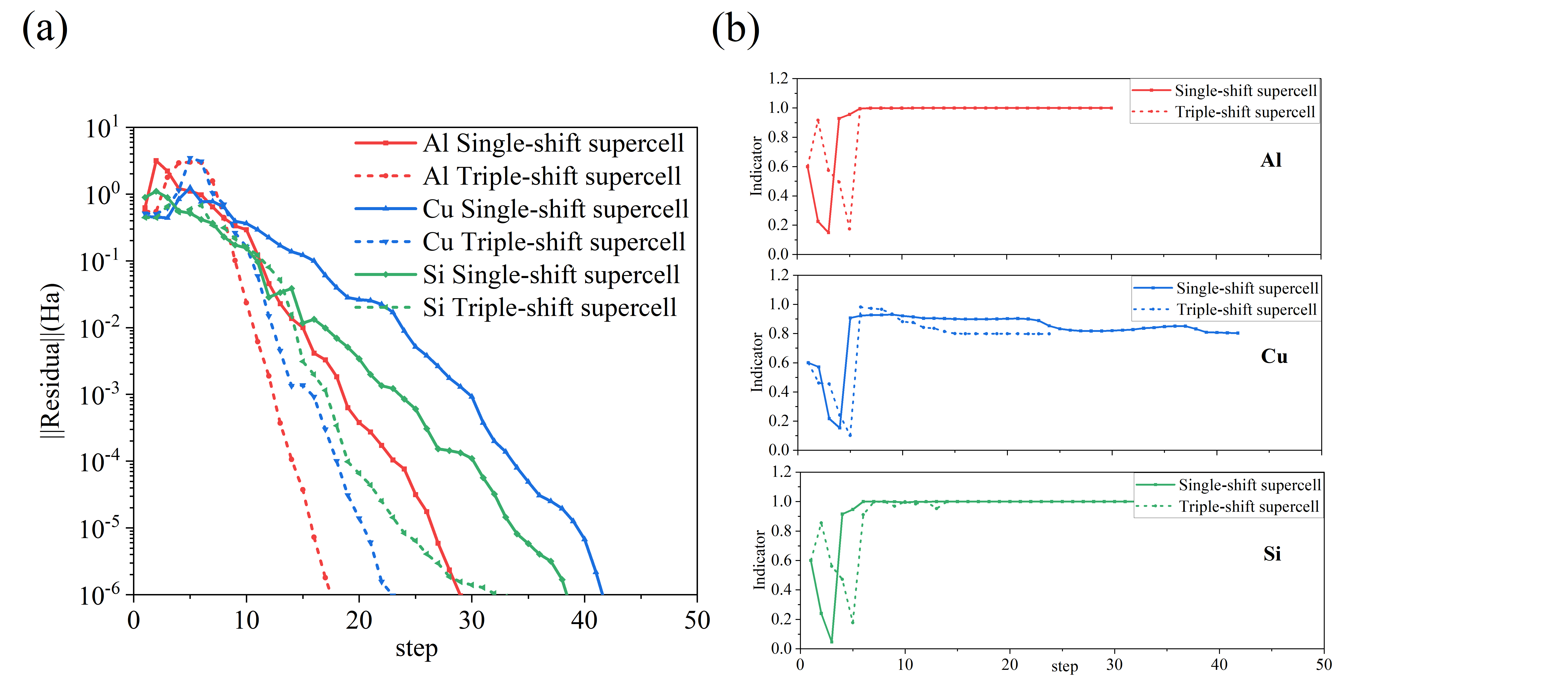}
	\centering
	\caption{\label{fig:7} (a) and (b) The convergence of residual and the indicator for Al, Cu and Si models of the single-shift and triple-shift supercell models by using the adaptive preconditioning scheme.}
\end{figure*}

We tested the convergence and posterior indicators of Al, Cu, and Si using the adaptive preconditioning scheme in the two supercell models. 
 We set $\lambda_{\rm Al}=0.6$, $\lambda_{\rm Cu}=0.8$ and $\lambda_{\rm Si}=0.6$. As shown in Fig.~\ref{fig:7}, small indicator appears around the 5 steps. Then the Kerker preconditioner is initiated, the indicators are approximately 1 and remains stable. It leads to a significantly improved convergence speed compared with the original algorithm. This scheme allows the system to adjust the choice of preconditioning method more reasonably without prior information.

\begin{figure}[h]
	\includegraphics[width=0.45\textwidth]{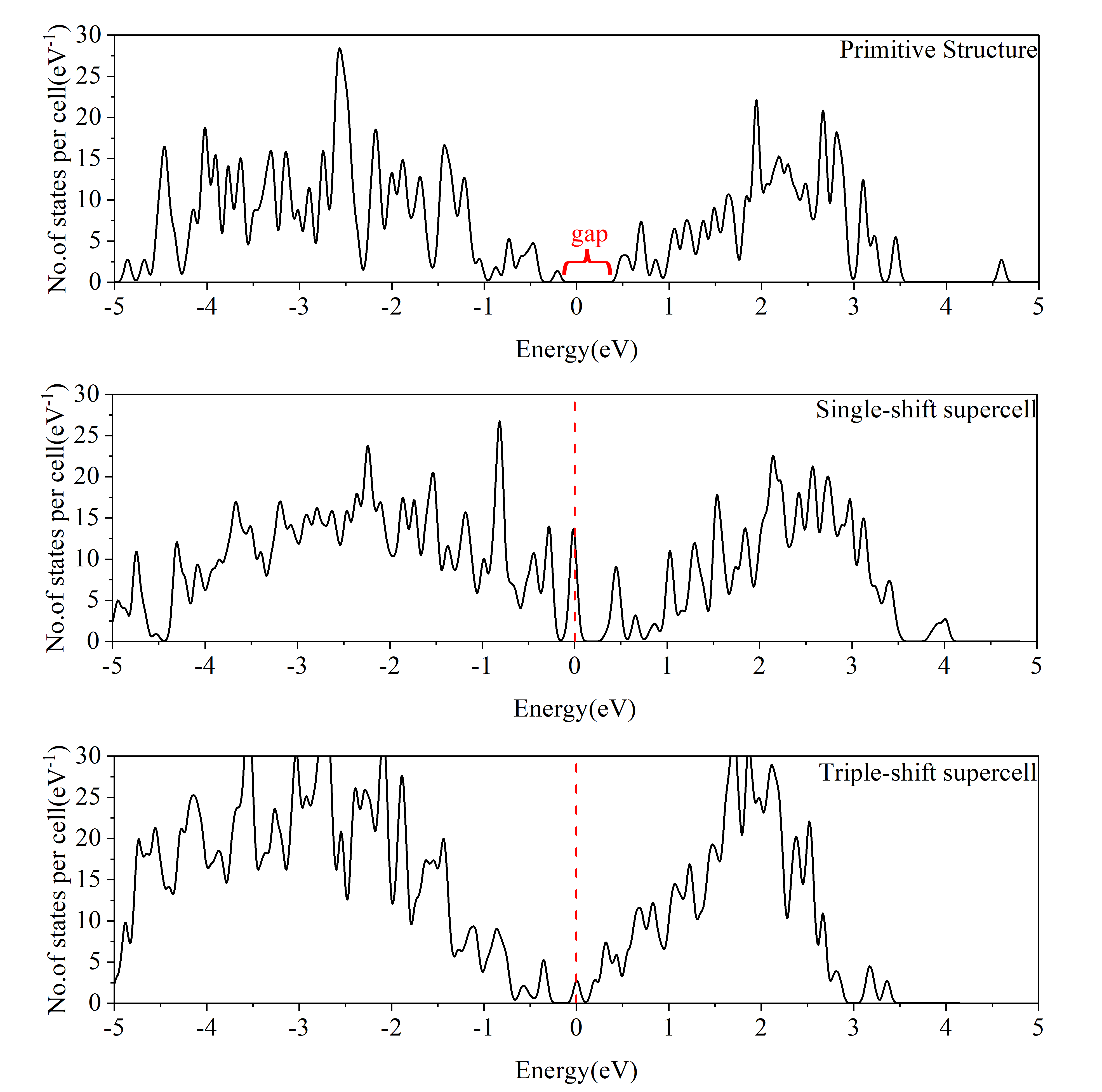}
	\caption{\label{fig:6} The DOS for the original structure, the single-shift supercell model, and the triple-shift supercell model of Si.}
\end{figure}

It is worth noting that the Kerker preconditioner is aimed at metallic systems, but in out tests, it is also effective for Si, which is a semiconductor. We calculated the density of states (DOS) for the original structure, the single-shift supercell model, and the triple-shift supercell model of Si, as shown in Fig.~\ref{fig:6}. Although the original structure of Si is insulated, in the single-shift and triple-shift supercell models, planar defects in stacking-fault structures generate free electrons, giving the system metallic properties.
Moreover, in the single-shift supercell model, the planar defects introduced by a vacuum layer enhanced the metallicity.
Our programs identified the long-wavelength divergence behavior of the Jacobian and initiated Kerker preconditioner improving convergence in these models. 
\section{Evaluation of GSFE curve calculation}\label{sec:4}
\begin{figure}[h]
	\includegraphics[width=0.45\textwidth]{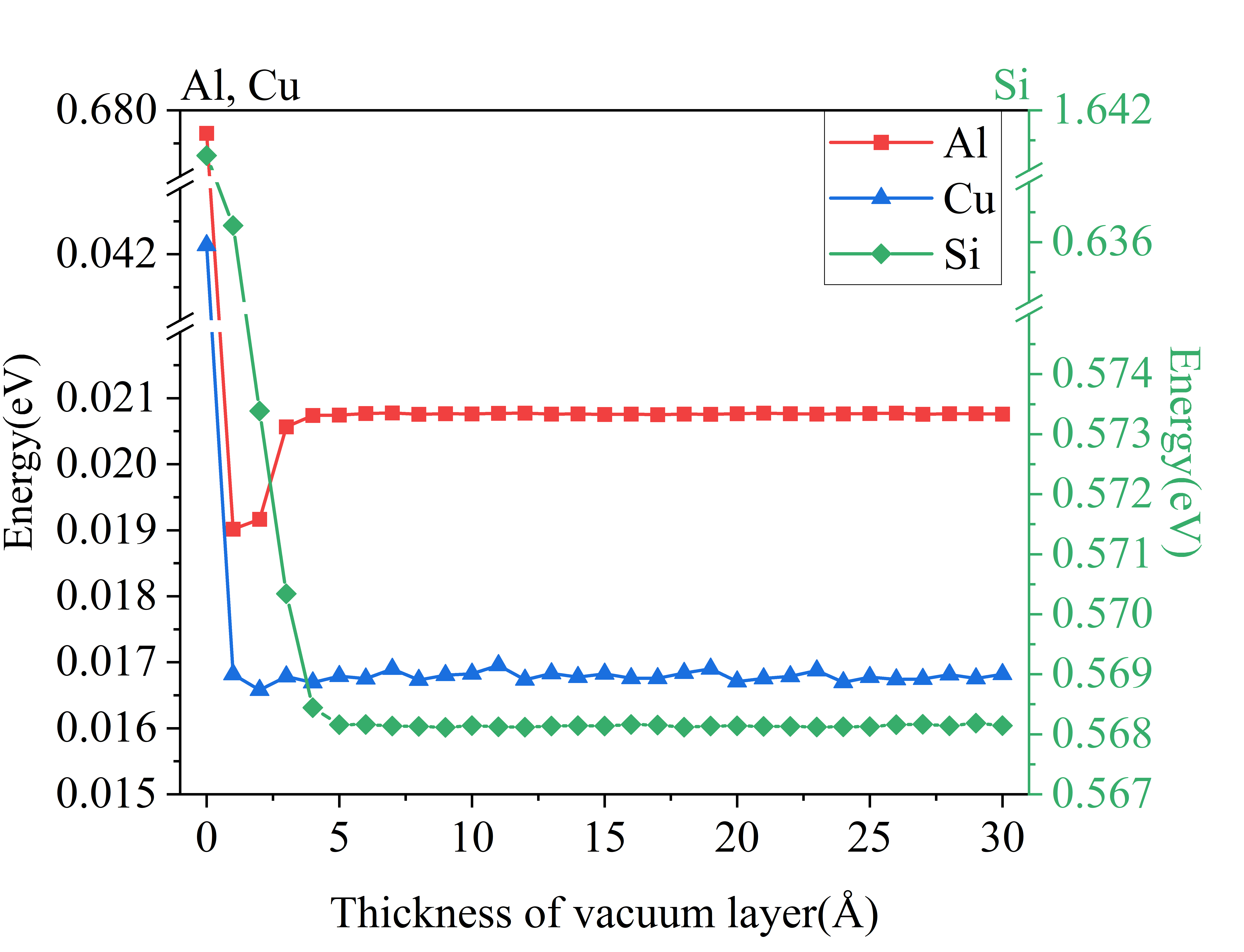}
	\caption{\label{fig:9} The energy of the single-shift supercell model changes with the thickness of the vacuum layer.}
\end{figure}

We first tested the effect of vacuum layers with different thicknesses on the energy for the single-shift supercell model. As shown in Fig.~\ref{fig:9}, the energies of all three materials converge in the range of $10 ^ {-4} $eV at 10 Å, indicating that they can prevent the interaction between the slabs and their periodic images. In addition, we also found that as the thickness of the vacuum layer increases, the convergence steps will gradually increase.

\begin{figure}
	\includegraphics[width=0.5\textwidth]{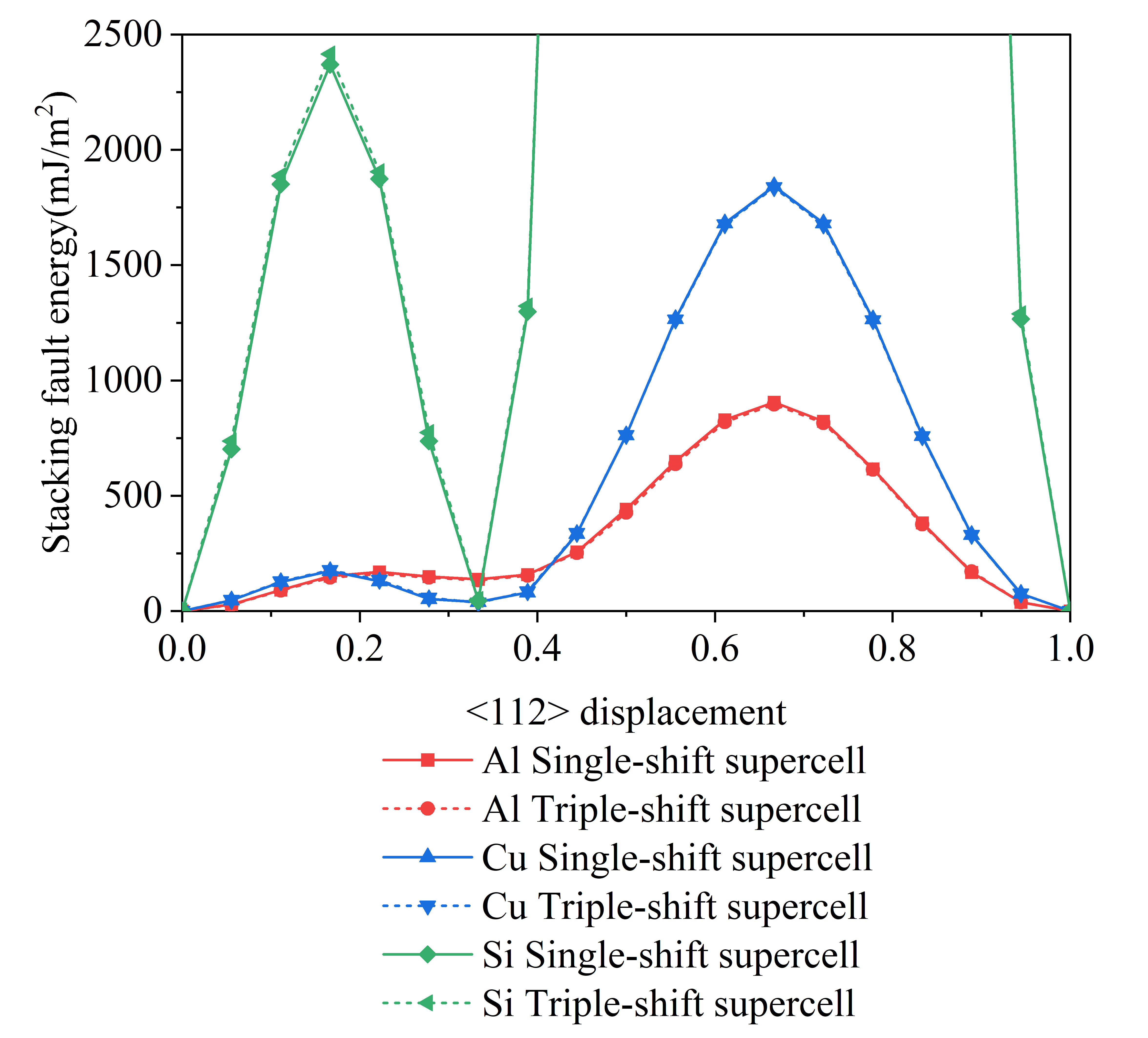}
	\caption{\label{fig:8} The GSFE curve of Al, Cu and Si for the single-shift and triple-shift supercell models.}
\end{figure}

Fig.~\ref{fig:8} shows the GSFE curves for Al, Cu, and Si of the single-shift and triple-shift supercell models.
We compared our intrinsic SFE values from other methods, and it is shown that the intrinsic SFE values of Al are between 120-166 $ mJ/m^2 $, of Cu, they fall within 35-45 $ mJ/m^2 $, and of Si, they are in 35-70 $ mJ/m^2 $. 
Our calculated results are within this range. 
The results of the single-shift and triple-shift supercell models are comparable.
And the shape of the GSFE curve is the same as the previous research, confirming the credibility of our computations.
\begin{table}[h]
	\caption{\label{tab:table1}
		Intrinsic SFE $ (mJ/m^2) $ for Al, Cu, and Si in different methods}
	\begin{ruledtabular}
		\begin{tabular}{ccccc}		
			&Method &Al &Cu &Si \\
			\hline
			&Sgl-shift& 138 & 39 & 45 \\
			&Tpl-shift& 134 & 36 & 50  \\
			&WIEN2K\cite{51}& 150 & 44 &    \\
			&MEAM\cite{54}& 150 & 43 &    \\
			&PAW& 112\cite{53},142\cite{wu2014generalized} & 36\cite{53} & 47\cite{55}   \\
			&EMTO\cite{52}& 107 & 47 &    \\
			&Tight-binding\cite{72}& 99 & 30 & \\
			& EXP.& 120-142\cite{han2008deformation},135\cite{smallman1970stacking},&35-45\cite{58}, & 50$\pm$15\cite{takeuchi1999stacking}   \\
			& &	120\cite{61},166\cite{15} &43.5$\pm $2\cite{1983903}&60$\pm$10\cite{foll1979direct}
	\end{tabular}
	\end{ruledtabular}
\end{table}

Then, We compared the computational cost for Al, Cu, and Si in the two models, as shown in Table~\ref{tab:table2}. Our CPU is Intel(R) Xeon(R) Gold 5218 CPU $ \textit{@} $ 2.30GHz; every system was tested with 32 cores in parallel.
The more atoms for the triple-shift supercell model produce the larger computation times than the single-shift supercell model in per electronic step, but the faster convergence results in similar computation times of the two models for SCF iteration calculations. In structural optimization, the triple-shift supercell models also have the fewer steps, leading to the less total time compared to the single-shift supercell model. 
Clearly, different defects yield distinct convergence behaviors, although Kerker preconditioner suppresses the long-wavelength divergence of the Jacobian, the planar defects introduced by a vacuum layer still slows convergence. 
In comparison, defects arising solely from stacking-faults have a milder impact than the planar defects introduced by a vacuum layer on convergence. 
\begin{table}[h]
	\caption{\label{tab:table2}
		Time-to-solution for Al, Cu and Si models of SFE calculations by using Kerker preconditioner}
		\small
	\begin{ruledtabular}
		\begin{tabular}{ccccccccc}
			&system &model &SCF-step  &1-SCF(t) \footnotemark[1]&relax-step&total(t) \footnotemark[2]\\
			\hline
			&Al& Sgl-shift&30  & 130.80	 & 28 & 16.26
			\\
			&& Tpl-shift&19 & 250.10 & 13 & 7.56 \\	
			&Cu& Sgl-shift&43  & 56.19  & 33 & 9.21\\
			&& Tpl-shift&24  & 116.96 & 11 & 3.98  \\
			&Si& Sgl-shift&41  & 1240.35 & 27 & 130.69  \\
			& & Tpl-shift&34 & 2129.94 &15  & 70.02  \\
		\end{tabular}
	\end{ruledtabular}
	\footnotetext[1]{Time unit: CPU seconds}
	\footnotetext[2]{Time unit: CPU hours}
\end{table}
\section{Conclusions}\label{sec:5}
Different supercell models for GSFE calculations introduce different degrees of defects, which affecting the convergence of both the SCF iteration and atomic relaxation. In the perspective of accuracy of GSFE calculations, the single-shift and triple-shift supercell models are comparable. However, the latter would be much more computationally efficient than the former since the degree of defect is diminished in the triple-shift supercell. In the absense of a priori knowledge, the adaptive preconditioning scheme proposed in this work can help to identify the underlying long-wavelength divergence behavior of the Jacobian and restore the convergence of the SCF iteration. This algorithm may be also applied to some heterojunctions such as metal-insulator contacts systems.

Our scheme is implemented in the FLAPW method, and enables the Elk-7.2.42 package to work out the GSFE much more efficiently compared to the original scheme. Obviously, the construction of the a posteriori indicator is independent of the basis functions, and the corresponding computational overhead is negligible. Therefore it may be easy to implement our scheme in other first-principles codes. The ability would faciliate high-precision computations of the characteristic quantities of the microstructure, which is significant for the simulation of the microstructure evolution.
\section{Declaration of Interest Statement}
The authors declare that they have no known competing financial interests or personal relationships that could have appeared to influence the work reported in this paper.
\section{Acknowledgment}
This work was supported by the National Natural Science Foundation of China (Grant Nos. 51925105, and U23A20537).
\bibliography{apssamp}

\end{document}